\documentstyle[11pt,paspconf,epsfig]{article}

\begin{document}

\title{The Observational Basis For Central Engines in Gamma-Ray Bursts}

\author{E.\ E.\ Fenimore and E.\ Ramirez-Ruiz\altaffilmark{1}}
\affil{MS D436, Los Alamos National Laboratory, Los Alamos, NM, 87545}


\altaffiltext{1}{Also Facultad de Ciencias, Universidad Nacional
Aut\'onoma de
M\'exico, Distrito Federal, M\'exico 04510}


\begin{abstract}

We review the observational differences between gamma-ray bursts occurring
on a single shell (such as in the external shock model) and multiple
shells (such as in the internal shock model).  The expected 
profile and
average spectral evolution from a single shell is compared to the average
of many bursts and found to be different.  The presence of gaps in many
gamma-ray bursts is a strong argument against a single shell because
an observer should see many causally disconnected regions at any one time.  
The rapid variability is also difficult to explain
from a
single shell because of the large number of causally disconnected regions.
The pulse width as a function of time in a burst should increase because
there is a one-to-one relationship between arrival time and the off-axis
angle of emission. The observations show that the pulse width does not
increase with time.  Finally, in GRB990123 there is evidence for
deceleration from the simultaneous optical observations, yet the
gamma-ray pulses show no lengthening of their pulse structure.  We
conclude that gamma-ray bursts are caused by a relatively small central
engine.

 \end{abstract}


\keywords{gamma-ray bursts, relativity}


\section{Introduction}

The rapid temporal evolution and GeV emission in gamma-ray bursts (GRBs)
indicate relativistic motion with bulk Lorentz factors of at least 100.
Two competing explanations have been suggested to explain the rapid time
variability.  In the ``external'' shock model
(M\'esz\'aros \& Rees 1993)
there is a very quick ($< 1$
s) release of energy at a central site that produces an expanding
relativistic shell.  That shell interacts over a long period of time
($10^6 - 10^7$ s) with the interstellar medium (ISM),  producing multiple
releases of
gamma rays.
The shell keeps up with the
photons it produces  in such a way that they are separated (spatially) by
a few
light-seconds even though they were emitted over a long period of time.  
These bursts of gamma rays arrive at the detector over a
modest range of times (10 - 100 s). 

The alternate theory is that the release of energy at the central site is
sporadic and lasts as long as we observe the burst to last. Each release
forms a shell which is closely related to an observed peak.  The gamma
rays might result from
shocks that occur when one shell runs into another and, hence, these
models are often called ``internal'' shocks
(Rees \& M\'esz\'aros 1994).

In this review, we present analyses based on causality and kinematics that
strongly argue that the GRB phase occurs in a small region and not on a
single shell.
We will establish six scenarios that cover all kinematically allowed ways
for a single shell to produce the gamma-ray time history. 
Observational evidence will rule out all six.  We conclude the source must
be smaller, as in the internal shock model.  We do not argue against
external shocks or for internal
shocks, {\it per se}, but rather against a single shell and for a central
engine that is small enough to
produce the typical GRB time history without violating causality or other
kinematic limits.

Given the large Lorentz factor, $\Gamma$, of the shell, one only sees
photons from the portion of the shell that is within angles 
$\sim \pm\Gamma^{-1}$ about the line of sight.  Thus, the shell must
effectively be aimed directly
at the observer.  In such a situation, the temporal variations in an
observer's
detector do {\it not}
tell the observer what time scale the source varied  in the {\it observer
rest frame}.  Time in a detector is not the detector's rest frame time. 
Rest frame time must be measured by clocks placed at all
sites in the rest frame.  Rather, a detector measures when the photons
arrive at a single location.  Normally this distinction plays no role, but
it does when the source is moving towards the observer.  This is not due
to
Lorentz transformations but due to superluminal effects.  Consider a
photon which is emitted by the shell as the shell leaves the central site.
If another photon is emitted $t$ sec later, it is behind the first photon
by
only $(c-v)t = ct/(2\Gamma^2)$ where $v$ is the velocity of the shell,
$\beta=v/c$, and $\Gamma=(1-\beta^2)^{-1/2}$.
These two photons will arrive at the detector separated in time by
$T=t/(2\Gamma^2)$.  (We denote the detector's rest frame time with $t$ and
the arrival time with $T$.) In contrast, clocks moving with the shell
would measure time $t'$ which is related to the observer's rest frame time
by the Lorentz factor: $t=\Gamma t'$.  So, if a shell expands in the
detector rest frame for $10^7$ s and $\Gamma = 200$, the detector sees
photons for 125 s while a clock on the shell would see emission for 25,000
s.

Most results in this paper originate from the observation that the
average
profile of GRBs tend to have a fast rise and slower decay.  The relatively
fast rise ($<$20\% of the duration) says that the shell is active only over
a short period.  As a result, the profile is dominated by curvature and
there is a one-to-one relationship between the originating angle of
the emission and the time of arrival at the detector.

The curvature of the shell is actually more important than the expansion.
The curvature is extremely small, but the expansion is contracted by
$2\Gamma^2$.  Photons produced from a part of the shell that is at an
angle $\theta$ off the line of sight to the observer must travel an
additional distance: $ct(1-\cos\theta)$.  For an angle near $\theta =
\Gamma^{-1}$, this additional distance corresponds to a delay in arrival
time of $t/(2\Gamma^2)$, the same as the separation in time due to the
expansion.

\begin{figure}
 \centerline{\epsfig{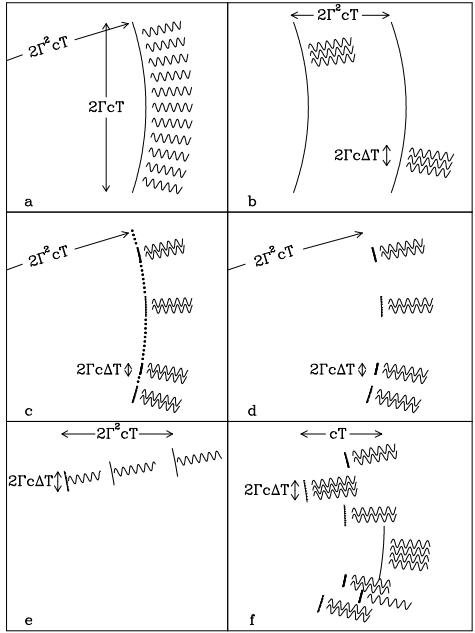}}
\caption{
Possible emission geometries for a single relativistic shell.
The observed average profile of GRBs has a sharp rise and slow decay which
is inconsistent with scenarios with full shells that emit over a wide
range of radii (b, f).  Gaps in GRBs are inconsistent with
scenarios that
have full shells and emit over a short range of radii (a, c, d).
The observed low filling factor indicates that most of the angular size
available to the shell does not emit.  This implies that the energy
reservoir for the burst must be $\sim1000$ times larger than estimated
from the
observed gamma-ray flux if the non-emitting regions (denoted by dots in c)
have bulk energy as do the regions that emit.  If the burst consists of
many fine jets, the reservoir does not have to be larger (see d).
Scenario (e) uses a narrow jet to get rid of curvature effects so it can
more easily have gaps and the average profile.  However, to have constant
pulse width throughout the burst, it requires no deceleration.
GRB990123 shows deceleration during the burst,
arguing against scenario (e).
}
\label{scenariofig}
\end{figure}

The six possible emission geometries from a single shell
are shown in Figure~\ref{scenariofig}. For example, the first
(Fig.~\ref{scenariofig}a) is the classic single shell that coasts for a
while and then turns on.  
Figure~\ref{scenariofig} will be explained as we discuss various
observations.

\section{Expected Emission from a Single Shell}

Assume the shell
expands for a time ($t_e$) in a photon quiet phase and then emits
uniformly for
a short period of time
 (i.e., scenario [a] in Fig.\ \ref{scenariofig}). That is, the production
of
photons is $P(t) = P_0\delta(t-t_e)$. 
 In terms of arrival time, the
{\it on-axis} emission will arrive at  $T_e = t_e/(2\Gamma^2)$ and the
off-axis emission will arrive later.  The relationship between the angle
on the shell and the arrival time depends only on the time since the
shell left the central site (if $\Gamma$ is constant, Fenimore,
Madras, \& Nayakshin 1996):
\begin{equation}
2\Gamma^2(1-\beta\cos\theta)=
(T/T_e)~~.  
\label{ANGLETIME}
\end{equation}
This equation establishes
a one-to-one relationship between the angle at which photons originate
and the time at which they 
 {\it arrive} at
the detector.  
It is from this equation that we determine how the
observed time history should evolve. The observed temporal variability
($V$) is
found by integrating over the volume where
photons arrive at the
detector at the same time and including relativistic beaming. For the
$\delta$-function production of photons in time (and radius), the
expected shape is (Fenimore, et al.\
1996):
\begin{eqnarray}
V_{\delta}(T,T_e) &  =  0~~~~~~~~~~~~~ & ~{\rm if} ~T<T_e\;,
\label{VSINGLE}
 \\
  & =   \psi T_e~\bigg({T \over T_e}\bigg)^{-\alpha-1}   & ~{\rm if}~T>T_e
\nonumber
\end{eqnarray}
where we have assumed that the rest frame photon-number
spectrum is isotropic with a power law with index $-\alpha$, and
$\psi$ is a constant.
This envelope is similar to a ``FRED'' 
(fast rise, exponential decay)
where  the shape of the slow, power law decay depends only on
the time that the shell expands before it emits ($T_e$).
The decay phase is due to
photons delayed by the curvature.
If GRBs had this shape, we could determine a best-fit $T_e$.   As a rule
of thumb, $T_e \sim 5 T_{FWHM}$
where $T_{FWHM}$ is the full width at half maximum of the profile
(Fenimore et al.\ 1996).

The expected spectral variation can also be found.
 GRB spectra
can often be fit by the so-called ``Band'' model (Band et al. 1993)
which consists of two power laws and the peak of the $\nu F_{\nu}$
distribution,
If $E_p^{\prime}$ is the peak of $\nu F_{\nu}$ in the rest frame of the
shell, the observed
$E_p$ is Doppler boosted.
The delayed photons are  boosted less  because
they originate from regions moving at the angle $\theta$ relative to
the on-axis regions.
 The Doppler boost as
a function of arrival time is
\begin{equation}
B_\delta(T,T_e) = \big[\Gamma(1-\beta\cos\theta)\big]^{-1}=
{1\over 2\Gamma}\bigg({T\over T_e}\bigg)^{-1}~~.
\label{BOOST}
\end{equation}
This equation  establishes
a one-to-one relationship between Doppler boost 
and the time at which they 
{\it arrive} at
the detector (Fenimore et al.\ 1996).

More complex envelopes can be found from weighted sums of
$V_\delta(T,T_e)$.
  For example, the shell might emit for a range of times during which it
collides with something.
We model  $P(T_e,T_0)$ to be non-zero  from $t_0$ 
to $t_{\rm max}$, and we assume
$P(T_e,T_0)$ can be approximated as a power law:
\begin{equation}
P(T_e,T_0)= P_0(T_e/T_0)^{-\eta}.
\label{PTT}
\end{equation}
 In terms of arrival time, the
{\it on-axis} emission will arrive between  $T_0 = t_0/(2\Gamma^2)$ and
$T_{\rm max} = t_{\rm max}/(2\Gamma^2)$.
 The
expected envelope, $V(T)$, is:
\begin{equation}
V(T)
 = \int_{T_{0}}^{T}
V_{\delta}(T,T_e)P(T_e,T_0) \,dT_e~~.
\label{FUNENVELOP}
\end{equation}
Due to the curvature of the shell, off-axis photons will be delayed, and  most 
emission will arrive later:
\begin{eqnarray}
V(T) & = 0   & {\rm
if}~~T < T_{0}
\label{ENVELOPE}
\\
  & =
{\psi P_0 \over \omega T_0^{-\eta}} 
{T^{\omega} - T_{0}^{\omega} \over T^{\alpha+1}}
 & {\rm if}~T_{0} < T < T_{\rm max}
\nonumber \\
  & =
{\psi P_0 \over \omega T_0^{-\eta}}
{T_{\rm max}^{\omega} - T_{0}^{\omega} \over  T^{\alpha+1}}
 & {\rm if}~T > T_{\rm max}
\nonumber
\end{eqnarray}
where $\omega=\alpha+3-\eta$ (Fenimore et al.\ 1996).

The envelope in equation (\ref{ENVELOPE}) is  similar to equation
(\ref{VSINGLE}), that is,  a ``FRED''
where the  rise depends mostly on the duration of the photon
active phase
($T_{\rm max}-T_0$) and the slow, power law decay depends mostly on
the final overall size of the shell
($T_{\rm max}$).

\section{Average Profile and Spectral Evolution}

The predicted profile of GRBs (Eq.~[\ref{ENVELOPE}]) and spectral
evolution
(Eq.~[\ref{BOOST}]) can be compared to the average profile and average
spectral evolution of GRBs (Fenimore  1999).  The average profile can be
found by scaling the duration of each burst
by a constant before they are
averaged.  This can be viewed as an ``aligned $T_{<{\rm Dur}>}$''
average in contrast to the ``aligned peak'' averages, such as those used by
Mitrofanov, Litvak, \& Ushakov (1997).  In the aligned peak average, each
burst
contributes to the average by aligning the largest peak.  The time
scale of the peak is conserved as it contributes to the average.
In the aligned $T_{<{\rm Dur}>}$ average, each burst contributes to the
average by
aligning the midpoint of the burst, and the time scale of the burst is
adjusted to a standard duration which we call $T_{<{\rm Dur}>}$.

The Burst and Transient Source Experiment (BATSE)  catalog
provides durations called $T_{90}$ and $T_{50}$ (Meegan et al. 1996).
For example,
$T_{90}$ is the duration which contains 90\% of the counts.  It is
defined by finding the duration that excludes the first 5\% and last
5\% of the counts in the burst.  There is a similar definition for $T_{50}$.
We estimate an average duration, $T_{<{\rm Dur}>}$, from $T_{90}$ and
$T_{50}$.  To first order,
$T_{<{\rm Dur}>}$ is $T_{90}/0.9$ or $T_{50}/0.5$.
By definition, the beginning
point for $T_{90}$ or $T_{50}$ must be at a point at which the count rate
is increasing. 
Thus, if we stretched each burst to a standard duration by scaling the
time by some multiple of $T_{90}$, there would be a coherent peak at the
first 5\% point and at the last 5\% point.
Rather, we define $T_{<{\rm Dur}>}$ to be a combination of
$T_{90}$ and $T_{50}$
to break up the coherency.  Specifically, we define
\begin{equation}
T_{<{\rm Dur}>} = {(T_{90} + T_{50})/2 \over 0.7}~~.
\label{TONE}
\end{equation}
\begin{figure}[t]
 \centerline{\epsfig{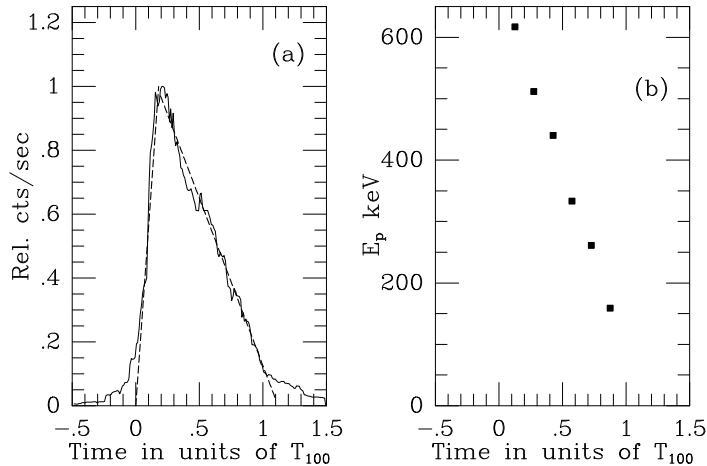}}
\caption{
The average temporal and spectral evolution of bright events with
intermediate durations ($T_{90}$ between 16 and 40 sec). (a) The average
time history.  The decay phase starting 20\% after the
beginning of the $T_{<{\rm Dur}>}$ period is
inconsistent with exponential decays and
power law decays.  Instead, the decay is consistent with a linear
slope.
(b) The average spectral evolution.  The spectral evolution is found by
fixing the low energy and high energy slopes at the average for the
bursts and allowing only the peak of the $\nu F_\nu$ distribution to vary.
The peak energy is also a linear function.
Thus, on average, the intensity is a
linear function of the peak of the $\nu F_{\nu}$ distribution.  This
temporal and spectral evolution
is inconsistent with that expected from a single shell.
From Fenimore (1999).
}
\label{t100fig}
\end{figure}
In Figure~\ref{t100fig}  we used the BATSE CONT data to investigate the
average spectral
and temporal evolution. 
The BATSE CONT data has 16 energy channels and 2.048 s time
resolution.
We used the 32 BATSE 3B events with $T_{90}$ between 16 and 40 s since
those
events have a sufficient number of CONT samples to investigate the
spectral evolution..
The resulting average time history appears to rise to a peak and then fall
linearly.  We have fit a variety of temporal shapes to the decay portion
between 20\% after the beginning of the $T_{<{\rm Dur}>}$ period to the end
of the $T_{<{\rm Dur}>}$ period.
We fit a linear function, an exponential function, and power law decays.
We particularly checked if a $T^{-1.4}$ power law would fit because that
type of decay is seen later during the x-ray afterglows (Piro, Matt, \&
Ricci 1997).
The linear fit was the best fit (Fenimore 1999). The power law and
exponential fit had
$\chi^2$ values that were 2.2 and 3.8, respectively, times larger and
they disagreed with the observations in a systematic way, failing to
agree with the observations at the ends of the time range.  A power
law with an index of -1.4 had a $\chi^2$ that was 6.4 times larger
than the linear fit.
The best linear function is:
\begin{eqnarray}
I &= 5.56 {T \over T_{<{\rm Dur}>}}~~~~~~~~~~~~&~~
{\rm if}~~ T < 0.18 T_{<{\rm Dur}>}
\label{AVEPROFILEB}
\\
{}~&=1.19 -1.06 {T \over T_{<{\rm Dur}>}}~&~
{\rm if}~~ T > 0.18 T_{<{\rm Dur}>}~~.
\nonumber
\end{eqnarray}
We calculated aligned $T_{<{\rm Dur}>}$ averages for each of the 16 CONT
 energy channels. From these, 
 six spectra were formed, each  covering 15\% of the $T_{<{\rm Dur}>}$
range.
The first one started 5\% after the beginning of the $T_{<{\rm Dur}>}$
range, and the last one ended at 5\% before the end of the
$T_{<{\rm Dur}>}$ range.
For each of these spectra, we fit
the ``Band'' spectral shape (Band et al., 1993).
This shape consists of a low energy slope ($\alpha$), the peak of the
 $\nu F_{\nu}$ distribution ($E_p$),
and a high energy slope ($\beta$).  We first fit the Band shape to the
sum of all six spectra.  
To investigate the average spectral evolution, we analyzed each of the
six spectra separately, fixing $\alpha$ and $\beta$ to their average
value.
Thus, the only free parameter is $E_p$.
 Figure \ref{t100fig}b shows the resulting spectral evolution.
 It is
a remarkably straight line:
\begin{equation}
E_p = 680 -600{T \over T_{<{\rm Dur}>}}~~~ {\rm keV}~~.
\label{AVENERGY}
\end{equation}

The decay phase of the average GRB profile argues against scenarios that
generate photons from a small range of radii (i.\ e., a, c, d in Fig.\
\ref{scenariofig}).  
The expected profile from a small range of radii should have a
power law decay phase (i.e., $T^{-\alpha-1}$, where $\alpha$  $\sim
1.5$) and the  peak of $\nu F_\nu$  should evolve as $T^{-1}$ (see eqs.\
[\ref{BOOST},\ref{ENVELOPE}]).  The observations indicate linear decays.

The rise phase of the average GRB profile argues against scenarios that
generate photons over a large range of radii (i.\ e., b, f).  
For example, in  scenario (b), a small portion of the shell  
emits at one
radius (small portion to produce the observed small $\Delta T$'s) and then
moves
to
another radius where another small portion emits. 
The shell has such activity for a long time, $t_{\rm
max}-t_0$, but the shell keeps up with
the photons it produces.  The resulting duration in arrival time is
$T_{\rm max}-T_0 = (t_{\rm max}-t_0)/(2\Gamma^2)$.  The average profile
should reflect the increasing surface area as the shell moves, that is,
equation (\ref{ENVELOPE}) gives a long rise and short decay rather than
the observed short rise and long decay.

In fact, expansion should not eliminate curvature effects,  even in the
extreme case of $T_0 = 0$ in equation (\ref{ENVELOPE}).
Dermer \& Mitman
(1999) have modeled gamma-ray generation by small clouds spread out over a
wide range of radii, effectively scenario (b).
 Their profiles tend to have longer rises than
observed  because $T_{\rm max}-T_0$ in equation (\ref{ENVELOPE}) is large.
The pulses at the end of their bursts are noticeably
longer due to residue curvature effects (and deceleration).
If they had
restricted their clouds to a smaller range of radii, the profile would
have a faster rise, but then the curvature effects would be stronger.

Scenario (f) also emits over a range of radii although much shorter than
scenario (b).  Scenario (b) emits at different radii {\it and} different
times.
The shell keeps up with the photons it produces so the range of time is
contracted by $2\Gamma^2$.  In scenario (f), the shell is fragmented,
so
exists at different radii at the same time.  The fragments can emit at
different radii but at nearly the same time.  The photons are born with
the separation that produces the observed duration. So, the range of radii
in scenario (f) is $cT$, much shorter than in scenario (b).  The fragments
in scenario (f)
appear to be independent of each other and there is no clear reason
why the resulting average profile would have a short rise or long decay.

\section{Filling Factor}

The rapid variability seen in GRBs implies small spatial structures.  The
few number of peaks in GRBs implies only a few such structures.  Thus,
only a fraction of the surface of the shell becomes active.  We define the
``surface filling factor'' to be that  fraction. (Fenimore et al.\ 1996,
1999b).  Let $A_N$ be the
area of an entity and $N_N$ be
the number of entities that (randomly) become active during the interval
$T_{\rm obs}$.  If $A_{\rm obs}$ is the area of the shell that can contribute
during $T_{\rm obs}$, then the surface filling factor is
\begin{equation}
f = N_N{A_N \over A_{\rm obs}}~= ~~N_N{A_N \over \eta
A_S}~
\label{EFF}
\end{equation}
where $\eta$ is the fraction of the visible area of the shell ($A_S$)
that
contributes during the interval $T_{\rm obs}$.

The number of entities can be determined from the observed fluctuations
in the time history. Variations about the overall envelope are due to a
combination of the Poisson variations in the number of emitting entities
and
the Poisson variations associated with the count statistics.
Assume for the moment that the contribution due to the count statistics
are small such that the variations come from the Poisson variations in
the number of contributing entities.
We first remove the envelope by fitting a polynomial function to it.
The observations are divided by the polynomial function,
so the result is a flat
envelope with variations due to the number of entities that, on
average, are active simultaneously.  The rate of occurrence of the
entities, $\mu_N$, can be found directly from the variations because the
observed mean level is $K\mu_N$ where $K$ is some constant and the
variance is $K^2\mu_N$. We define $N$ to be the observed mean level of
the flattened envelope
and $\delta N$ to be the root mean square of the flattened envelope.
In the case of no contribution from the counting statistics, 
$\mu_N$ would be  $(N/\delta N)^2$.
Here, we have implicitly assumed that all entities are identical.
This is supported
by the fact that peaks within GRBs usually are similar to each other.
The actual root mean square of the flattened envelope is a combination
of the variance due to the entities and the variance due to the counting
statistics, $\sigma^2_{CS}$. We assume they add in quadrature, that is, 
$N^2/(\delta N)^2 = \mu_N + \sigma_{CS}^2$.  
We estimate $\sigma_{CS}$ to be root mean square of many Monte Carlo
realizations of the flattened envelope.
The rate of occurrence of entities is
\begin{equation}
\mu_N = {N^2 \over (\delta N)^2} - \sigma_{CS}^2~~.
\label{ENTITYRATE}
\end{equation}
This rate is the number of events per the time scale of the entities.  
Thus, the total number of 
entities that occur
within a period $T_{\rm obs}$ is
\begin{equation}
N_N = \mu_N{T_{\rm obs} \over \Delta T_p}
\label{ENTITYNUM}
\end{equation}
where $\Delta T_p$ is the time scale for a single entity.

We have
consider several processes that relate the size of an entity causing
a peak to an observed peak duration, $\Delta T_p$.
Here, we consider the two most
likely processes for the formation of a peak:  regions that grow and 
regions formed by the interaction with the ISM.

Consider a gamma-ray producing region that grows at a speed
near that of light, $c_s$, for a period $\Delta t^\prime$ in the rest
frame of the
shell where presumably the region is symmetric.  This growth might be
associated with a developing shock.  Let $\Delta
r^\prime_\parallel$ be the radius of the region in the rest frame along the
direction of the motion and $\Delta r^\prime_\perp$ be the radius in the
perpendicular direction such that
\begin{equation}
\Delta r^\prime_\parallel = \Delta r^\prime_\perp = c_s\Delta t^\prime
= c_s \Gamma^{-1} \Delta t~~.
\label{LORENTZ}
\end{equation}
The sizes in the rest frame of the detector are related to the sizes in
the rest frame of the shell as: $\Delta r_\parallel =
\Gamma^{-1}\Delta r^\prime_\parallel$ and 
$\Delta r_\perp = \Delta r^\prime_\perp$.

Combining the effects of the movement of the shell during the growth
with the maximum size that the entity can grow in time $\Delta t$, we
find that the duration in arrival time is 
\begin{equation}
\Delta T_p = {\Delta t \over \Gamma^2}
\bigg[\big({1 \over 2}\big)^2 + \big({c_s \over c}\big)^2\bigg]^{1/2}~.
\label{DTGROWTH}
\end{equation}
The size of the emitting entity is
\begin{equation}
A_N = \pi \Delta r^2_\perp = \pi
\big({c_s \Delta t \over \Gamma}\big)^2 =
{\pi c_s^2 \Gamma^2 \Delta T_p^2 \over
\big[({1 \over 2})^2 + ({c_s \over c})^2\big]}~~.
\label{SIZEA}
\end{equation}

The alternative cause of a peak resulting from the shell is that the shell
interacts with an ambient object such as an ISM cloud.  Presumably, the
object is symmetric such that $\Delta R_{\perp} = \Delta R_{\parallel}
= \Delta R_{\rm amb}$. (We use lower case $\Delta r$ for an object
that grows in a shell and upper case $\Delta R$ for an ambient
object.)
The contribution to the peak duration from the
time the shell  takes to move through the cloud (i.e., $\Delta R_{\rm
amb}/[c\Gamma^2]$) is negligible compared to the time the shell takes 
to engage the perpendicular size of the object. 
This engagement time is caused by the curvature of the shell.
At an angle $\theta$
from the line of sight, the time to engage the object is 
$\Delta T_{\Delta R_{\perp}} = \theta\Delta R_{\rm amb}/(2c)$.  Note
that Table 2
of Fenimore et al.\ (1996) was incorrect for $\Delta
T_{\Delta R_{\perp}}$; see Sari \& Piran (1997).
At a typical angle of $\theta \sim \Gamma^{-1}$,
\begin{equation}
\Delta R_{\rm amb}  = {c\Delta T_p  \Gamma \over 2}
\label{RAMB}
\end{equation}
and
\begin{equation}
A_N = \pi \Delta R^2_{\rm amb} = {\pi c^2 \Gamma^2\Delta T_p^2 \over 4}~~.
\label{SIZEB}
\end{equation}
Thus, both the case of shocks that grow from a seed and the case of 
shells running into
ambient objects are similar. If $c_s = c/3$,  then $A_N$ from equation
(\ref{SIZEA}) is 16/13 times larger than from equation (\ref{SIZEB}).
These two scenarios only differ  by a constant the order of unity.

A common misconception is that one can just use an ISM cloud that covers
most of the shell's surface.  Only the instantaneous
interaction between two {\it plain parallel} surfaces oriented
perpendicularly
to our line of sight can produce a short
peak from large surfaces.  A curved surface has sources limited in size to
that of equation (\ref{SIZEB}).

\begin{figure}
 \centerline{\epsfig{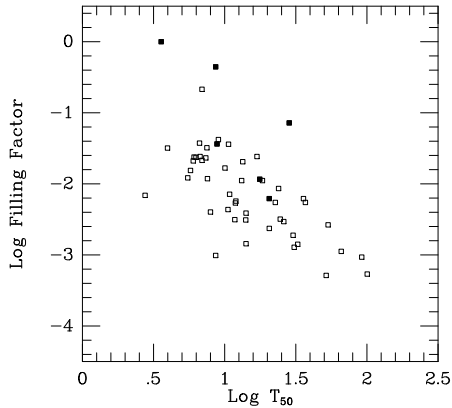}}
\caption{
Typical values of the fraction of a relativistic shell that becomes
active during a GRB as a function of the duration of the emission
($T_{50}$).
The six solid squares are FRED-like BATSE bursts for
which direct estimates of the size of the shell can be made.  The 46 open
squares are long complex BATSE bursts where we estimate the size in a manner
similar to the FRED-like estimate.
Under most conditions,
the efficiency is $\sim 0.1 \Delta T/T$.
These low values imply that either only a small fraction of
the shell converts its energy into gamma-rays or that GRBs consist
of very fine jets with angular sizes much smaller than
$\Gamma^{-1}$.
From Fenimore et al.\ (1999b).
 }
\label{fillingfig}
\end{figure}

The final ingredient for the calculation of the surface filling factor
is the area
of the shell visible to the observer:
\begin{equation}
A_S = 2\pi R^2(T)(1-\cos\theta_{\max})~,
\label{AREASHELL}
\end{equation}
where $\theta_{\max}$ is either the angular width of the shell  or it is
$\sim \Gamma^{-1}$, whichever
is smaller. Before the shell starts to
decelerate, presumably $\theta_{\max}$ is larger than
$\Gamma_0^{-1}$. Using $R(T) = 2\Gamma^2_0 c T$,
\begin{equation}
A_S = \pi \big[\Gamma_0^{-1}R(t)\big]^2 = 4\pi c^2\Gamma_0^2 T^2~~.
\label{AREACONST}
\end{equation}
Thus, during the constant $\Gamma$ phase,
the filling factor is
\begin{equation}
f =  N_N \big[{\Delta T_p \over T}\big]^2 {1 \over k\eta}=
   \big[{N^2 \over (\delta N)^2}-\sigma^2_{CS}\big] {\Delta T_p \over
k\eta T}
\label{EFFEQ}
\end{equation}
where $k$ is 16 for ambient objects and 13 for entities that grow from a
seed (see differences between equations (\ref{SIZEA}) and (\ref{SIZEB}).

From the expected envelope (eq.\ \ref{ENVELOPE}), $\eta$ is 1 when
$T=0.8T_{50}$ (Fenimore et al.\ 1999b). Thus, the filling factor can be
found from $T_{50}$, the mean and variance of the GRB time history, Monte
Carlo estimates of $\sigma_{CS}^2$, and the typical widths of individual
pulses  such as found by Norris et al.\ (1996).
In Figure \ref{fillingfig} we show the distribution of surface filling
factors
as a function of
burst duration $T_{50}$. The solid squares are the FRED-like bursts, and
the open squares are the long complex bursts.  Although some of the smooth
FRED-like bursts can have surface filling factors near unity,
most bursts have
values on the order of $5 \times 10^{-3}$.

Small filling factor implies two things. First, that one should see many
peaks in GRBs, the order of $(T/\Delta T)^2 \sim 10^4$.
Second, if the filling factor is small because there are many more places
on the shell that do not convert their bulk energy to gamma rays, then the
energy reservoir needs to be $f^{-1}$ times larger (Fenimore et al.\ 1996,
Sari \& Piran 1997).

Consider scenarios  (c) and
(d) in Figure \ref{scenariofig}.  
In both, 
only a a single radius becomes active, as consistent with the observed
fast rise of the average GRB profile. In both, 
only a
fraction of the surface becomes gamma-ray active.   In scenario (c), 
only patches on the shell become active so the inter-patch regions
(denoted by the dotted lines in Fig.\ \ref{scenariofig}c) do not convert
their kinetic
energy to gamma-rays.  Scenario (d) is very similar except the inter-patch
regions are devoid of material.  Effectively, scenario (d) is many narrow
jets.  Small filling factor raises the required energy reservoir only in
scenario (c) but not (d).  We estimate the energy requirement by observing
the
gamma-ray flux at earth and making corrections (i) for the distance
to the object, (ii)
for emission not seen because it beamed away from us, and (iii) for the
efficiency of converting bulk energy into gamma-rays. Corrections (i) and
(ii) are the same for all scenarios.  Correction (iii) will always include
correcting for the microscopic physics of generating gamma rays.  Small
filling factor implies a macroscopic issue: are there portions of the
shell that never (or barely) converts its bulk energy? Most of the bulk
energy in scenario (c) is never converted to gamma rays so it requires
a reservoir that is $f^{-1}$ time larger than for scenario (d).  

In other scenarios (e.\ g., b, f), the  low filling factor also implies a
few emitting entities relative
to the available surface area. If that is accomplished by having the rest 
of the surface area not convert its energy,
the energy requirement will be larger by $f^{-1}$.  If it is accomplished
by many fine jets, the energy reservoir needs not be larger.

We consider it unlikely that most
of the shell never produces gamma-rays, especially if the process involves
forming an external shock in the ISM.  The ISM should decelerate all
portions of a shell.  However, the alternative (many fine jets in scenario
[d]) is unlikely from hydrodynamic considerations  (Sari \& Piran 1997).
When the opening angle of a jet is less than $\Gamma^{-1}$, the jet will
expand preventing such fine jets.  Thus, a low filling factor (with or
without requiring a larger energy reservoir) presents a strong case
against scenarios b, c, d, and f.

\section{Gaps}

\begin{figure}
 \centerline{\epsfig{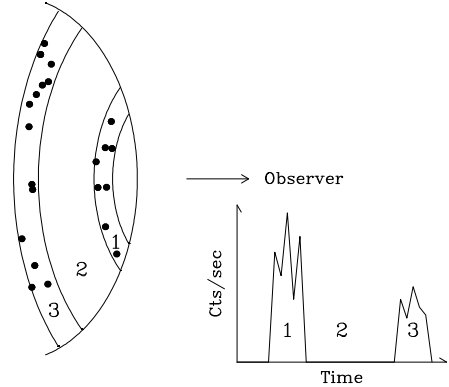}}
\caption{
Schematic of the relationship between the emission on a shell and
the observed time history.
The
curvature delays the photons from off-axis regions such that at any
one time, the observer sees photons from an annulus oriented around
the line of sight.  The
perpendicular size of the shell is $\sim 2c\Gamma T$ whereas a causally
connected entity (represented by the dots) is only $2c\Gamma\Delta T$.
Here, $T$ and $\Delta T$
are the time in the time history and a typical time scale of variation.
Gaps imply that large causally disconnected regions
do not emit (e.g., region 2 produces gap 2 in the time history).
The number of entities in each annulus determines the
variability of the time history.  The ``filling factor'' is the fraction
of the shell occupied by the emitting entities and is typically $10^{-3}$.
}
\label{gapfig}
\end{figure}

Gaps or precursors in GRBs produce one of the strongest arguments against
a single
relativistic shell.
The sharp rise in the average profile (see section 3) indicates that the
shell emits for
a short
period of time (i.e., $t_0$ to $t_{\rm max}$ in eq.\ [\ref{ENVELOPE}] is
short
relative to the duration of the event), so that the
shape of the overall envelope is dominated by photons delayed by the
curvature.
During the decay phase, the  one-to-one relationship between time of
arrival and angle (eq. [\ref{ANGLETIME}]) means that, at any one time,
only  an annulus oriented
about the line of sight contributes photons to the observer. 
 Gaps in the time history indicate that some annuli
emit while others do not (see Fig.
\ref{gapfig}).  From equation (\ref{SIZEA}) we know that the maximum
radius
of a causally connected region is $2c\Gamma\Delta T$ whereas the radius of
the visible shell is $2c\Gamma T$ (cf.\ eq.\ \ref{AREACONST}). Thus, for
gaps to occur, a large, causally disconnected annulus about the line of
sight must coordinate its emission.  This is the strongest argument
against scenarios consistent with the rapid rise of the average profile,
that is, those that use a full shell emitting over a short range of radii
(scenarios a, c, d in Fig. \ref{scenariofig}).

\section{Constant Pulse Width}

\begin{figure}
 \centerline{\epsfig{file=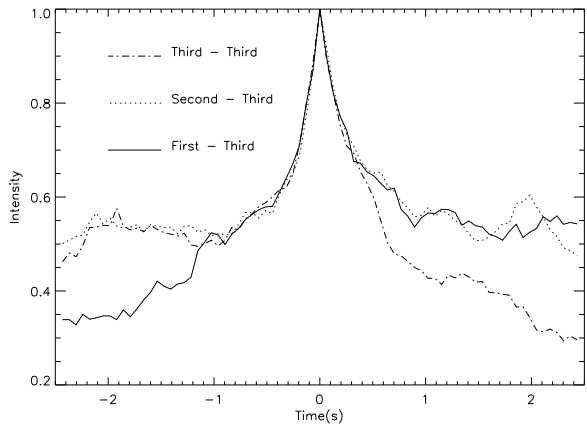,height=6cm,clip=}}
\caption{Average peak alignment from 53 bright BATSE bursts with
durations longer than 20 s.The three curves show the   
average pulse shape for the largest peak in the first third, second
third, and last third of the bursts. We find no  significant   
change, during the gamma-ray phase, in the average peak width over at
least 2/3 of $T_{90}$.
From Ramirez-Ruiz \& Fenimore (1999).
}
\label{constantwid}
\end{figure}

A visual inspection of the BATSE catalog of multiple-peaked time
histories reveals that peaks usually have about the same duration at
the beginning of the burst as near the end of the burst. 
The aligned peak method measures the average pulse temporal
structure, each burst contributes to the average by aligning the
largest peak (Mitrofanov 1997). 
To characterize the average evolution of peak widths with time,
we used all 53 bursts from
the BATSE 4B Catalog that were longer than 20s and brighter than 5
photons s$^{-1}$ cm$^{-2}$. Each burst was required to have at least one
peak, as
determined by a peak-finding algorithm (similar to Li \& Fenimore 1996), in
each  third of its duration. The largest peak in each third was
normalized to unity and
shifted in time, bringing the largest peaks of all bursts into
common alignment. This method was applied in each third
of the duration of the bursts. Thus, we  obtained one curve
of the averaged pulse shape for each third of
the bursts (as shown in Figure \ref{constantwid}). The average profile is
notably
identical in each 1/3 of $T_{90}$, the difference in widths is less than
1\%.

  There are two sources of pulse width from a shell: angular 
effects and deceleration.  
Assuming that the pulse duration in the rest frame of the shell is not 
dependent on the direction to the observer, the pulse width at two times A 
and B are related as
\begin{equation}
{\Delta T_{\rm B} \over \Delta T_{\rm A}} =
{\Gamma_{\rm B} (1-\beta\cos\theta_{\rm B}) \over
\Gamma_{\rm A} (1-\beta\cos\theta_{\rm A})}
\label{WIDRELATE}
\end{equation}
where $\theta_{\rm B}$ and $\theta_{\rm A}$ are the angles
responsible for the
emission.
From equation (\ref{ANGLETIME}),
\begin{equation}
\Gamma(1-\beta\cos\theta_{\rm A}) = {T_{\rm A} \over 2\Gamma T_0}
\label{TLAM}
\end{equation}
where $T$ must be measured from when the shell left the central
site
and $T_0$ is the time of the peak of the emission.  Before the shell
decelerates, $\Gamma(T)$ is constant and pulse widths ought to scale at
$T_B/T_A$.  Thus, the lack of pulse width evolution in Figure
\ref{constantwid} is strong evidence against scenarios (c) and (d) where
there is a strong relationship between angle and time.

Scenario (f) attempts to get around the constant peak width constraint by
upsetting the one-to-one relationship between arrival time and angle.   If
$\Gamma$ varies on an angular scale much
smaller than $\Gamma^{-1}$, the shell could fragment. After coasting for a
while, the fragments would be spread out in time commensurate with the
observed duration of GRBs.  This requires variations in $\Gamma$ of at
least a factor of $\sqrt{2}$ (Fenimore, et al.\ 1996, 1999b). The peaks
arrive at the detector arranged by the value of $\Gamma$ of the fragment
that produce the emission.  Although we cannot specify what process
actually makes the peaks, it is surprising that the resulting peaks have
very consistent pulse width yet were formed from a systematic trend in
$\Gamma$

The only way to be consistent with gaps, low filling factor, short rise
time in the average profile, and constant peak width is scenario (e),
that is, a very
narrow jet (to eliminate curvature effects and allow a low filling factor)
with no deceleration (to be
consistent with constant pulse width).

\section{GRB990123}

On January 23, 1999, the Robotic Optical Transient Search Experiment
(ROTSE) discovered strong optical emission (9th mag) during a gamma-ray
burst (Akerlof et al.\ 1999). This
behavior was
predicted a few weeks before (Sari \& Piran 1999a).
The optical emission peaked at about 45 s after the trigger and is very
consistent with the emission expected from forward and reverse external
shocks that form when the shell decelerates in the ISM (Sari \& Piran
1999b).  This provides an excellent opportunity to test scenario (e) in
Figure~\ref{scenariofig}.  Scenario (e) is consistent with constant
pulse widths, gaps, the average profile, and small filling factor {\it if}
the
shell is not decelerating. 
Since there is gamma-ray emission after the
optical peak in GRB990123, we can test it for a widening of the pulses.
If the
gamma-ray are being made on the single shell responsible for the optical
emission, we ought to see the pulses widen due to the deceleration.

After deceleration started, during the afterglow, $\Gamma(T) \propto
T^{-3/8}$ so the combined 
effects of angle and deceleration at a typical angle of $\theta =
\Gamma^{-1}$ gives from equation (\ref{WIDRELATE}):
\begin{equation}
{\Delta T_{\rm B} \over \Delta T_{\rm A}} =
\bigg[{T_{\rm B} \over T_{\rm A}}\bigg]^{11/8}
~~.
\label{PULRAT}
\end{equation}
If the shell is relatively narrow (range of $\theta < \Gamma^{-1}$) as in 
scenario (e),
then the pulses only grow due to the deceleration:
\begin{equation}
{\Delta T_{\rm B} \over \Delta T_{\rm A}} =
\bigg[{T_{\rm B} \over T_{\rm A}}\bigg]^{3/8}
~~.
\label{PULRATNAR}
\end{equation}
We analyzed four periods after the peak of the optical emission in a
manner similar to Figure~\ref{constantwid} (Fenimore, Ramirez-Ruiz, Wu 
1999a). 
It is likely that the shell started to leave the central site at about 
the time when the first gamma rays were emitted.  BATSE detected emission 
quite early in this burst, so we will use the BATSE time for $T$. The
first time we analyzed is at $\sim 45$ s after the start and last time is
at $\sim 82$ s. 
Based 
on this, we expect the pulse widths to increase by about a factor of 2.3 
(eq. [\ref{PULRAT}]) if the shell is wider than $\Gamma^{-1}$, and 1.25 (eq.
[\ref{PULRATNAR}]) if the shell is much narrower than $\Gamma^{-1}$. We
found that 
$\Delta T_B/\Delta T_A$ was $1.034\pm0.035$. The minimum expected from
just deceleration was rejected at the $6\sigma$ level.  Thus, we conclude
that scenario (e) can be rejected as well: the pulses do not get wider
when the shell is decelerating.

\section{Summary}

   GRBs time histories display two salient features: long durations and
chaotic variations.  We have presented six scenarios for how a single
relativistic shell produces these features.  These scenarios differ
primarily in how they explain {\it duration}.  Some produce the long
duration by delays caused by the curvature of the shell (a, c, d in Fig.\
\ref{scenariofig}). Curvature dominates when the shell emits effectively
at one radius and one time.   Scenarios (a, c, d) differ in how they
produce the chaotic variations. Duration associated with expansion
(scenarios b, e) are caused by a shell that emits at different radii and
different times.  Scenario (b) still has some curvature effects while (e)
emits over a very narrow angle.  The scenario (f) shell has lost its
curvature  due to different parts of the shell having different
speeds. The parts spread out and emit at about the same time,
but at radii that are separated by distances that light can travel during
a typical GRB duration.

A shell that coasts and then emits over a very short range of radii (and
therefore, time) will have a duration set by the curvature.  Indeed, the
average GRB profile
has a sharp rise and long decay characteristic of a profile made by
curvature.  However, the details of the decay do not match the
observations, the decay is linear whereas power laws were expected.  That
difference is probably not large enough to reject curvature as the source
for the duration.  Gaps provide the most potent
argument against models
that rely on curvature.  Gaps indicate that some large causally
disconnected
regions in an annulus about the observer's line of sight emit while others
do not.  This point alone should eliminate
scenarios (a, c, d).

The second problem with curvature is that the one-to-one relationship of
arrival time with angle on the shell implies that pulses should get wider
later in the burst because the Lorentz factor changes.  This is not
observed.

Scenario (b) uses expansion to produce the duration.  Expansion has a
phase when the available surface area (i.\ e., within $\Gamma^{-1}$) grows
with time and, therefore, the expected average profile should have a long
rise  ($T_{\rm max} >> T_0$ in eq.\ \ref{ENVELOPE}).  The observed fast
rise in the average profile eliminates this scenario.  In addition, even
in an extreme case ($T_0 = 0$), the curvature effects are comparable to
the expansion effects. One should still see some increase in the pulse
width during the burst and none is seen.  Scenario (e) avoids both
problems by being very narrow: no increase in surface area and no
curvature to contend with.  To be consistent with constant pulse width
would require no deceleration, but that is what is observed in GRB990123.

Scenario (f) avoids curvature effects by breaking up the shell  and avoids
expansion effects by using spatial distances to get the burst duration.
There is no apparent reason why it would produce the average burst
profile.

The large number of causally disconnected regions while we see
relatively few peaks in GRBs leads to low ``filling factor''. Only about
$10^{-3}$ of the available surface becomes active.  This is a strong  
argument against scenarios (c) where the non-emitting regions of the shell
have bulk energy that is never converted to gamma-rays.  Scenario (d)
eliminates those regions by being many fine jets.  Scenarios (b, f) would
also either require much
more energy or be made up of many fine jets to accommodate the low filling
factor.

Our scenarios span the combinations of emission at constant radii,
constant time, varying radii, and varying time except one:
roughly constant
radius and varying time.  Since the radius of a single shell is set by the
time ($R=vt$),
that combination is not possible with a single shell.  Rather, constant
radius with varying time is a central engine:  the central source provides
the duration of the event through the multiple releases of energy over a
time commensurate with typical burst durations.  We conclude that the
observations strongly argue that a single shell is not responsible for the
gamma-ray phase. Rather, a central engine is required to explain
the duration and chaotic nature of GRBs.  

\acknowledgments
This work done under the auspices of the US Department of Energy.

%
%

\end{document}